\documentclass[journal]{IEEEtran}
\def\baselinestretch{1.2}

\usepackage{tikz}
\graphicspath{{images/}}
\usepackage{amsmath}
\usepackage{latexsym, amssymb}
\usepackage{graphicx}
\usepackage{amsmath, amsbsy}
\usepackage{amsopn, amstext}
\usepackage{ifpdf,hyperref}
\usepackage{cancel, color}
\usepackage{epstopdf}

\DeclareMathOperator{\Col}{Col}

\DeclareMathOperator{\lcm}{lcm}

\def\cal{\mathcal}

\def\ra{\rightarrow}

\def\d{\delta}

\def\D{\Delta}

\def\0{{\bf 0}}

\newcommand{\R}{{\mathbb R}}

\newtheorem{thm}{Theorem}[section]
\newtheorem{dfn}[thm]{Definition}
\newtheorem{prp}[thm]{Proposition}

\newtheorem{cor}[thm]{Corollary}
\newtheorem{rem}[thm]{Remark}

\begin{document}

\title{A Comprehensive Survey on STP Approach to Finite Games}

\author{Daizhan Cheng$~^{\dagger}$,  Yuhu Wu $~^{\ddagger}$, Guodong Zhao$~^*$,  Shihua Fu$~^{**}$
	\thanks{This work is supported partly by the National Natural Science Foundation of China (NSFC) under Grants 62073315, 61074114, and 61273013.}
	\thanks{$~^{\ddagger}$: Key Laboratory of Intelligent Control and Optimization for Industrial Equipment of Ministry of Education, Dalian University of Technology, P.R. China (e-mail: wuyuhu@dlut.edu.cn).}
    \thanks{$~^{*}$: School of Mathematics and Statistics, Shandong Normal University, Jinan 250014, P.R. China  (e-mail: zgd\_qufu@126.com).}
  \thanks{$~^{**}$: School of Mathematical Sciences, Liaocheng University, Liaocheng, Shandong, P.R. China (e-mail: fush\_shanda@163.com).}
    \thanks{$~^{\dagger}$: Key Laboratory of Systems and Control, Academy of Mathematics and Systems Sciences, Chinese Academy of Sciences,
		Beijing 100190, P. R. China (e-mail: dcheng@iss.ac.cn; Tel.: +86 10 82541232), Corresponding author.}
}


\maketitle

\begin{abstract}
Nowadays the semi-tensor product (STP) approach to finite games has become a promising new direction. This paper provides a comprehensive survey on this prosperous field. After a brief introduction for STP and finite (networked) games, a description for the principle and fundamental technique of STP approach to finite games is presented. Then several problems and recent results about theory and applications of finite games via STP are presented. A brief comment about the potential use of STP to artificial intelligence is also proposed.
\end{abstract}


\IEEEpeerreviewmaketitle

\section{Introduction}

As twin branches of applied mathematics, game theory and control theory were born in the forties of last century. Both of them were stimulated partly by  World War II. Wiener's book \cite{wie48} was considered as the starting
symbol of control theory.  Similarly to control theory, von Neumann and Morgenstern's book \cite{von44} is the starting symbol of game theory. These two disciplines are aimed to manipulate certain objects to reach humans' goals. The major difference between them lies in the objects. The control objects are usually machines, while the playing opponents of a player in games are intelligent players, who are able to ``anti-control" you.

Roughly speaking, game theory can be divided into two broad categories: non-cooperative games and cooperative games. The equilibrium, which is proposed by Nash and then named after him,  is the key issue in non-cooperative game \cite{nas51}. One of the pioneers in cooperative game is Shapley. A famous imputation in cooperative game has been named by Shapley value \cite{gal62}. Both Nash and Shapley are the winners of the Nobel Prize in economics in 1994 and 2012 respectively.

Recently, there is a rapidly growing interest in the application of game-theoretic concepts and tools to control.
The reason for this trend was pointed by L. Guo as follows \cite{guo11}: ``When facing some intelligent objects such as intelligent machines and intelligent networks etc., the existing control theory may not be applied directly.
Because the controllers and the controlled objects may have game-like interactions. Putting certain game factors into a control framework is an important research topic, which is inevitable in dealing with some social and economical problems. Meanwhile, combining game factors can tremendously enlarge the development of control theory and its applications."  Some inter-relationship between them was demonstrated by \cite{bas99}, which also sets up an excellent example for applying control technique to game theory.

Nowadays, because of the development of computer science, artificial intelligence (AI) becomes more and more active in science and technology as well as in our daily life. In addition to computer-based numerical techniques, both the game theory and control theory also play an important role in the development of AI.

\begin{figure}
\centering
\setlength{\unitlength}{4.6 mm}
\begin{picture}(19,9)\thicklines
\put(0,1){\framebox(3,1.5){$Logical\atop System$}}
\put(4,1){\framebox(3,1.5){$Control \atop Theory$}}
\put(8,1){\framebox(3,1.5){$Game\atop Theory$}}
\put(12,1){\framebox(3,1.5){$Computer \atop Science$}}
\put(16,1){\framebox(3,1.5){$Numerical \atop Method$}}
\put(1.8,3.5){\framebox(3.4,1.5){$Hybrid \atop System$}}
\put(5.8,3.5){\framebox(3.4,1.5){$Game-based\atop Control$}}
\put(13.8,3.5){\framebox(3.4,1.5){$Learning\atop Algorithm$}}
\put(3,6.5){\framebox(13,1.5){$Artificial~~Intelligence$}}
\put(2.5,2.5){\vector(0,1){1}}
\put(4.5,2.5){\vector(0,1){1}}
\put(6.5,2.5){\vector(0,1){1}}
\put(8.5,2.5){\vector(0,1){1}}
\put(14.5,2.5){\vector(0,1){1}}
\put(16.5,2.5){\vector(0,1){1}}
\put(3.5,5){\vector(0,1){1.5}}
\put(7.5,5){\vector(0,1){1.5}}
\put(15.5,5){\vector(0,1){1.5}}
\put(3.6,5.5){Modeling}
\put(7.6,5.5){Control}
\put(15.6,5.5){Realization}
\end{picture}
\caption{Architecture of AI \label{Fig.1.1}}
\end{figure}
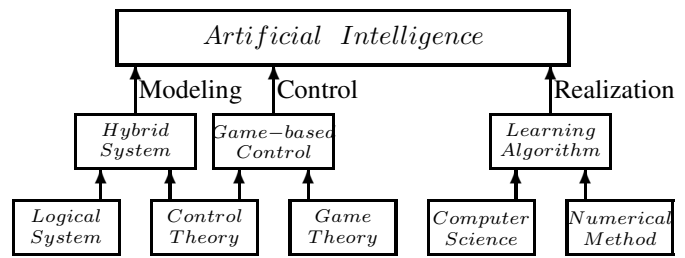

 In the authors' personal opinion, which is partly motivated by the argument in \cite{zha20}, the framework of AI can be depicted by Fig. \ref{Fig.1.1}. There the combination of logical system theory with control theory yields hybrid systems (including cyber-physical systems); the combination of control theory with game theory yields
 game-theoretic control \cite{mar09}; the combination of computer science with numerical method yields learning algorithm. Finally, hybrid system theory can be used for modeling AI; Game-based control is used for controlling AI, and learning algorithm is used to realize the control of AI.

It is our anticipation that game theory will play a key role in AI in the future.

Recently, the semi-tensor product (STP) of matrices has been used to finite game theory, and a series of new results have been achieved. It has been proved that STP is a convenient and powerful tool in dealing with many problems of finite games. The purpose of this paper is to explain the principle of this new approach, and then give a comprehensive survey on the results obtained by this approach.

The outline of this paper is as follows. After this introduction, Section 2 includes a brief introduction of finite games and  STP. Then an explanation is provided to answer ``why STP is suitable for modeling and analyzing finite games?" Section 3 presents the main results obtained for finite game theory using STP. Section 4 is a brief concluding remark.

\section{Finite Game via STP}

\subsection{Finite Game}

\begin{dfn}\label{d2.1.1} \cite{gib92} A finite non-cooperative game can be described as a triple $G=(N, {\cal S}, c)$, where
\begin{itemize}
\item[(i)] $N=\{1,2,\cdots,n\}$ is the set of players;
\item[(ii)]
$${\cal S}_i={\cal D}_{k_i}
$$
is the set of strategies (or actions) of player $i$,
where
$
{\cal D}_k:=\{1,2,\cdots,k\},
$
and $i=1,\cdots,n.$
$$
{\cal S}=\prod_{i=1}^n {\cal S}_i
$$
is called the profile set of the game;

\item[(iii)]
\begin{align}\label{1.1.1}
c_j:~{\cal S}\ra \R
\end{align}
is the payoff (or utility) function for  player $j$, $j=1,\cdots,n$ respectively, and
$c:=\left\{c_1,\cdots,c_n\right\}$.
\end{itemize}
\end{dfn}

\begin{dfn}\label{d2.1.2} In a finite game $G=(N,S,c)$, a profile
$$
s=(x^*_1,\cdots,x^*_n)\in {\cal S}
$$
is called a Nash equilibrium if
\begin{align}\label{1.1.2}
\begin{array}{r}
c_j(x_1^*,\cdots,,x_j^*,\cdots,x_n^*)\geq c_j(x_1^*,\cdots,x_j,\cdots,x_n^*)\\
\forall ~j=1,\cdots,n.
\end{array}
\end{align}
\end{dfn}

The physical meaning of a Nash equilibrium lies in that no player has anything to gain by a unilateral change of strategy if the strategies of the others remain unchanged. Hence, they have no incentive to change their own strategies unilaterally.

\begin{dfn}\label{d2.1.3} \cite{bil00} A finite cooperative game is a 2-tuple $G=(N,v)$, where
\begin{itemize}
\item[(i)] $N=\{1,2,\cdots,n\}$ is the set of players;
\item[(ii)] $v:2^N\ra \R$ is called the characteristic function.
\end{itemize}
\end{dfn}

A subset $S\subset N$ (equivalently, $S\in 2^N$) is called a colleague. In a cooperative game we are particularly interested in what a colleague can gain in a game. Hence, the characteristic function represents the achievement of each colleague.

The main purpose of a cooperative game is to find a fair (reasonable) distribution for each individual. This is called an imputation. An imputation is a function $\pi: N\ra \R$, which represents the payoff for each player.

\subsection{STP of Matrices}

\begin{dfn}\label{d1.2.1}\cite{che11,che12}
Let $A\in {\cal M}_{m\times n}$ and $B\in {\cal M}_{p\times q}$.
Denote
$$
t:=\lcm(n,p).
$$
Then  the STP   of $A$ and $B$ is defined as
\begin{align}\label{1.2.1}
A\ltimes B:=\left( A\otimes I_{t/n}\right)\left(B\otimes I_{t/p}\right)\in {\cal M}_{(mt/n) \times (qt/p)}.
\end{align}
\end{dfn}

\begin{rem}\label{r1.2.2}
\begin{itemize}
\item[(i)] It is a generalization of conventional matrix product.

\item[(ii)] It keeps all major properties of conventional matrix product available.

\end{itemize}
\end{rem}

Denote by $\D_k:=\Col(I_k)$ the set of columns of the identity matrix $I_k$, and $\d_k^j=\Col_j(I_k)$ the $j$-th column of $I_k$. A matrix $A\in {\cal M}_{m\times n}$ is called a logical matrix, if $\Col(A)\subset \D_m$. The set of logical matrices in ${\cal M}_{m\times n}$ is denoted by ${\cal L}_{m\times n}$. Hence, a logical matrix $A\in {\cal L}_{m\times n}$ can be expressed as $A=[\d_m^{i_1},\d_m^{i_2},\cdots,\d_m^{i_n}]$. A brief notation of $A$ is
$$
A=\d_m[i_1,i_2,\cdots,i_n].
$$

Let $f:{\cal D}_m\ra {\cal D}_n$ be a mapping from a finite set to a finite set. Then we can identify $j\in {\cal D}_m$ with its vector form $\vec{j}:=\d_{m}^j\in \D_{m}$. In this way, $f$ can be considered as a mapping $f: \D_m\ra \D_j$.

\begin{prp}\label{p1.2.3} Let $f:{\cal D}_m\ra {\cal D}_n$. Then there exists a unique matrix $M_f\in {\cal L}_{m\times n}$, such that as the argument is expressed into the vector form, we have
\begin{align}\label{1.2.2}
f(x)=M_fx.
\end{align}
\end{prp}

As a corollary,  Proposition \ref{p1.2.3} can be extended into a more general form.

\begin{cor}\label{c1.2.4} Let $X_i\in {\cal D}_{k_i}$, $i=1,2,\cdots,n$ and $Y_j\in {\cal D}_{p_j}$, $j=1,2,\cdots,m$.
Denote by $x_i=\vec{X}_i$ and $y_j=\vec{Y}_j$ the vector form of $X_i$ and $Y_j$ respectively, and $x=\ltimes_{i=1}^nx_i$, $y=\ltimes_{j=1}^my_j$. Let
$$
Y_j=f_j(X_1,X_2,\cdots,X_n),\quad j=1,2,\cdots,m.
$$
Then there exists a unique matrix $M_F$, called the structure matrix of the mapping $F=(f_1,f_2,\cdots,f_m)$,
such that
\begin{align}\label{1.2.3}
y=M_Fx,
\end{align}
where $M_F\in {\cal L}_{\rho\times \kappa}$ ($\rho=\prod_{j=1}^mp_j$, and $\kappa=\prod_{i=1}^nk_i$).
\end{cor}

\subsection{Formulation of Finite Games via STP}

Let ${\cal G}_{[n;k_1,k_2,\cdots,k_n]}$ be the set of non-cooperative games which have $|N|=n$, $|S_i|=k_i$, $i=1,2,\cdots,n$. Using the vector form of strategies, we have the following vector/matrix expressions:
\begin{itemize}
\item[(i)] Vector form of strategies:
\begin{align}\label{1.2.4}
S_i=\D_{k_i},\quad i=1,2,\cdots,n.
\end{align}
\item[(ii)] Vector form of payoff functions:
\begin{align}\label{1.2.5}
c_i=V^c_ix,\quad i=1,2,\cdots,n,
\end{align}
where the row vector $V^c_i\in \R^{\kappa}$ is the structure vector of payoff $i$, $x_i\in \D_{k_i}$ is the vector form of strategy of player $i$, and $x=\ltimes_{i=1}^nx_i$ is a profile.
\item[(iii)] Strategy profile dynamics:
if the game is an evolutionary game (i.e., repeated game), then the strategies of a player can be expressed as a function of previous strategies. Say, the evolution is Markov-type, that is,
\begin{align}\label{1.2.6}
x_i(t+1)=f_i(x_1(t),x_2(t),\cdots,x_n(t)),\nonumber\\
\quad i=1,2,\cdots,n,
\end{align}
then using Proposition \ref{p1.2.3} and  Corollary \ref{c1.2.4}, the evolution equation (\ref{1.2.6}) can be expressed as
\begin{align}\label{1.2.7}
x_i(t+1)=M_ix(t),\quad i=1,2,\cdots,n,
\end{align}
where $M_i\in {\cal L}_{k_i\times \kappa}$ is the structure matrix of $f_i$.
Furthermore, all the $n$ equations of (\ref{1.2.7}) can be equivalently expressed into a unified equation as
\begin{align}\label{1.2.8}
x(t+1)=Mx(t),
\end{align}
where $M=M_1*M_2*\cdots*M_n$ (here $*$ is the Khatri-Rao product) is the structure matrix of the overall strategy profile evolutional equation.
\end{itemize}

From the above argument, one sees the fundamental clue for formulating a non-cooperative game via STP.

Similar to non-cooperative game, cooperative game can also be formulated via STP.
Denote by ${\cal G}^c_n$ the set of cooperative games with $|N|=n$. Let $S\in 2^N$. Define the index function of $S$ as
${\cal I}_S:N\ra {\cal D}=\{0,1\}$ as
$$
{\cal I}(p)=
\begin{cases}
1,\quad p\in S,\\
0,\quad p\not\in S.
\end{cases}
$$
Then ${\cal I}_S\in {\cal B}^{n}$ is a Boolean vector. Say, ${\cal I}_S=(i_1,i_2,\cdots,i_n)$, then the vector form of $S$ is defined as
\begin{align}\label{1.2.9}
V_S=\d_{2^n}^i,
\end{align}
where
\begin{align}\label{1.2.10}
i=(1-i_1)2^{n-1}+(1-i_2)2^{n-2}+\cdots+(1-i_n)+1.
\end{align}
Then for a finite cooperative game $G=(N,v)\in {\cal G}^c_n$.
\begin{itemize}
\item Characteristic function:
There exists a unique row vector $U_v\in \R^{2^n}$ such that
\begin{align}\label{1.2.11}
v(S)=U_vV_S,\quad S\in 2^N.
\end{align}
\item Imputation:
Let $\phi(v)=(\phi_1(v),\phi_2(v),\cdots,\phi_n(v))$ be an imputation of $G=(N,v)$. Then there exists a matrix $M_{\phi}\in {\cal M}_{n\times 2^n}$, called the structure matrix of the imputation $\phi$, such that
\begin{align}\label{1.2.12}
\phi(v)=U_vM_{\phi}.
\end{align}
\end{itemize}

From the above argument, one sees that the matrix approach via STP is also efficient  for formulating a cooperative game.

\section{Main Results for Finite Games via STP}

\subsection{Networked Evolutionary Games}

\begin{dfn}\label{d3.1.1} A networked evolutionary game (NEG) can be described by a triple $G_e=((N,E), G, \Pi)$, where $(N,E)$ is the network graph; $G\in {\cal G}_{[2;k,k,\cdots,k]}$ is a symmetric game with two players, which is played by players $i$ and $j$, provided $(i,j)\in E$; $\Pi$ is the strategy updating rule (SUR), which determines how each player updates his strategy at the next step.
\end{dfn}

Note that in an NEG, the SUR should depend on local information only. The STP approach to NEGs was proposed in \cite{che12}. For an NEG with Markovian SURs, \cite{guo13} first provided an STP-based evolution equation for a special SUR, which has the form as (\ref{1.2.8}). Then \cite{che15} obtained the evolution equation for general SUR. Some important results such as (i) necessary and sufficient condition for stability strategies; (ii) evolutionary equivalence, etc. are also revealed. Consensus of NEGs was formulated and investigated in \cite{zha18b}. 

\subsection{Potential Games}

\begin{dfn}\label{d3.2.1} Consider a finite game $G=(N,S,C)$.
$G$ is a potential game (PG) if there exists a function $P:S\ra \R$, called the potential function, such that for $\forall$ $i\in N$ and  $\forall$ $s_{-i}\in S_{-i}$ and $\forall x,y\in S_i$, it holds that
\begin{align}\label{3.2.1}
\begin{array}{l}
c_i(x, s_{-i})-c_i(y,s_{-i})=
P(x, s_{-i})-P(y,s_{-i}),\\
\qquad i=1,\cdots,n.
\end{array}
\end{align}
\end{dfn}

A PG has some very nice properties such as (i) it has pure Nash equilibrium; (ii) it is equivalent to a congestion game, etc. \cite{mon96}. Later on, it becomes the kernel of
game-theoretic control \cite{gop11}.

However, verifying whether a finite game is a PG is a long-standing problem. As pointed by Hino \cite{hin11}: ``It is not easy, however, to verify whether a given game is a potential game." The investigation before us was through iteration, and the computational complexity was reduced from $O(k^4)$ \cite{mon96} to $O(k^3)$ \cite{hof02} to $O(k^2)$ \cite{hin11}. Using STP, \cite{che14} provided a closed form solution as follows: Construct a linear equation, called the potential equation. The main result is: a game is potential, if and only if, its potential equation has solutions. Following \cite{che14}, \cite{liu16} provided an algorithm based on potential equation and shown this algorithm reaches the minimum computational complexity.

Consider an NEG $G_e=((N,E), G, \Pi)$. It is interesting that when the fundamental network game $G$ is potential, no matter how the network graph $(N,E)$ is, the $G_e$ is also potential \cite{che14}. This fact is very useful in considering NEGs.

\subsection{Weighted Potential Games}

A weighted potential game (WPG) has the same nice properties as PG. So it might enlarge the PG applicable  sets of games significantly. \cite{wan20} proposed a more general WPG, called the coset-weighted PG, which is the most general WPG. \cite{wan17b} has extended the concept of WPG to weighted harmonic game.

A barrel in using WPGs to practical problems is, it is hard to verify whether a game is a WPG if the weights are unknown as in a real problem. Recently, \cite{chepr} provided a numerical method to solve the weighted potential equation and  weights simultaneously.

\subsection{Congestion	Games}

As a special class of non-cooperative games, congestion
games were first studied by Rosenthal \cite{Rosenthal73}, in which
multiple players compete for finite resources. By constructing a potential function, Monderer and Shapley \cite{mon96}
proved that each congestion game possesses at least one
pure-strategy Nash equilibrium. Due to this special property, congestion game has been widely applied
to  many practical problems, such as electric
vehicles charging \cite{Li2019new}, routes choice \cite{Zhou2020},  etc. Based on STP technique,  the existence of the weighted potential function for
 the congestion games with player-specific utility, and the congestion games with player-specific costs and resource
failures, are  proved, respectively in \cite{le19} and \cite{Wang2021}.

Congestion game theory is particularly suitable for traffic systems. Recently, using STP to formulate traffic congestion games, \cite{zha21b} proposed an optimal utility design technique. Another interesting result is about dynamics and convergence of hyper-networked evolutionary games with time delay in strategies \cite{zha21}.

\subsection{Vector Space Structure of Finite Games}

Consider a $G=(N,S,c)\in {\cal G}_{[n;k_1,k_2,\cdots,k_n]}$. Since
$$
c_i(x)=V^c_ix,
$$
we can construct a vector
$$
V^G:=[V^c_1,V^c_2,\cdots,V^c_n],
$$
then it is obvious that $G$ is completely determined by $V^G\in \R^{n\kappa}$. Hence ${\cal G}_{[n;k_1,k_2,\cdots,k_n]}$ has a natural vector space structure as
$$
{\cal G}_{[n;k_1,k_2,\cdots,k_n]}\sim \R^{n\kappa}.
$$
The decomposition of ${\cal G}_{[n;k_1,k_2,\cdots,k_n]}$ was first considered by some MIT scholars \cite{can11}. They presented the following orthogonal decomposition:

\begin{align}\label{3.4.1}
{\cal G}_{[n;k_1,\cdots,k_n]}=\rlap{$\underbrace{\phantom{\quad{\cal P}\quad\oplus\quad{\cal N}}}_{Potential\quad games}$}\quad{\cal P}\quad\oplus\quad
\overbrace{{\cal N}\quad\oplus\quad{\cal H}}^{Harmonic\quad games},
\end{align}
where ~${\cal P}$, ~${\cal N}$, ~${\cal H}$ are pure potential games, nonstrategic games, and pure harmonic games respectively.

This result is elegant and well known within game theory society. Unfortunately,  \cite{can11} used
algebraic topology as its fundamental tool and the decomposition result is based on a graphic decomposition theorem. Hence, it is not easily understandable. Moreover, the orthogonality is based on a weighted inner product. Using STP, \cite{che16} proved a similar decomposition based on linear algebra and the orthogonality with a standard Euclidean inner product.

This vector space and subspace structure is of significant importance, because in applications a game may not be potential, but it may be evolutionary equivalent to one of its ``neighborhood game"  that is potential. Such a game is called a near PG, which has similar properties as a PG and is useful in game-theoretic control.

\subsection{Symmetric/Antisymmetric Games}

Symmetric games are popular in the real world. For instance, prisoner's dilemma, rock-scissors-paper, etc. are symmetric.
The vector subspace structure of symmetric games was proposed in {\cite{che17}.

As an opponent type of games, antisymmetric games are considered in \cite{hao18}.  Then based on symmetry and antisymmetry, another orthogonal decomposition was discussed. Using STP, the following decomposition is obtained:

\begin{align}\label{3.5.1}
{\cal G}_{[n;k_1,\cdots,k_n]}={\cal S} \oplus {\cal A}\oplus {\cal K},
\end{align}
where ~${\cal S}$,~${\cal A}$,~${\cal K}$ are symmetric, skew-symmetric, and asymmetric games respectively.

\subsection{Relationship Between Potential and Symmetry}

A game is called a Boolean game if $|S_i|=2$, $i=1,2,\cdots,n$. For a Boolean game, \cite{che18} shows the following relationship between its potential and symmetry:
$$
\begin{array}{l}
\mbox{Boolean}+\mbox{Symmetric} \Rightarrow \mbox{Potential}\\
\mbox{Boolean}+\mbox{Potential} \not\Rightarrow \mbox{Symmetric}\\
\end{array}
$$
Some other interesting games and their properties were demonstrated.

\subsection{Strategy Stability}

The stability and stabilization of strategies of a finite game are  fundamental topics for evolutionary games. This problem has been discussed. The case when there is a time-delay has also been investigated. Certain necessary and sufficient conditions were obtained for both cases \cite{wan16,wan17}.

The stability of some specified NEGs is also investigated via STP. For instance, NEGs with time-varying network topology or with time-delays in updating were discussed in \cite{fu18b} and \cite{wan16,wan20b} respectively.

Stability to a mixed profile is theoretically interesting and practically important. In this case, the profile dynamics is a Markov process. This problem has been discussed in \cite{din17,li17}, certain necessary and sufficient conditions were obtained.

\subsection{Incomplete Profile Games}

Incomplete profile games exist widely in real world. For instance, in a chess game, there exists certain situation where some strategies are not allowed to use. In this case the Nash e equilibria and optimal strategies are state-depending. This kind of games has been discussed in \cite{zha18bb}. Furthermore, the PG under restricted profile is also discussed, and the method for verifying potential and finding potential function for incomplete profile games is obtained in \cite{pan20}.

\subsection{Games with Bankruptcy}

In an economic-based game, the strategy profiles may be forbidden for certain restrictions, which are named by bankruptcy. Taking bankruptcy into consideration as additional strategies and using STP, the evolution of such games were formulated, and their certain properties were investigated  in \cite{den19,fu17,fu18}.

\subsection{Some Special Types of Games}

The competitive diffusion game was discussed in \cite{li18}, method for verifying the existence of pure Nash equilibrium is proposed. Strategy consensus and uniform controllability  of NEG were investigated in \cite{li18b}. In \cite{zha18c} the long memory games were studied. The long memory games have been converted into time delay NEG systems. Regulation of games has been presented and investigated in \cite{din20}.

Using STP, the dynamics of NEGs with switched time delay has been modeled and investigated in \cite{zhe21}. Profile-dynamic based fictitious play was discussed in \cite{zhang21b}.

\subsection{Strategy Optimization}

Strategy optimization in man-machine games aims to optimize the humans' payoff, which may be considered as an optimization problem. STP approach was first proposed by \cite{zha11}. Based on STP, \cite{wan18} provided a systematic algebraic formulation and algorithm for searching optimal strategy of static games.  In public good games, the optimal strategies are aimed to maximize overall benefits. See \cite{fu20} for an STP approach. \cite{che20} provides a  sketch on optimization via game theoretic control.

\subsection{Hyper-networked Games}

Consider an NEG. If the network graph is a hyper-graph, then the game becomes a hyper-networked game. Potential hyper-networked game was considered in \cite{liu17}. Modelling and the stability of potential  hyper-networked were presented and investigated.

\subsection{Finite Element Approach to Continuous Games}

To apply the technique developed for STP approach to finite games to continuous games, a finite element method was proposed to convert a continuous game to a quantized finite game. The numerical algorithm is presented and the convergence of the Nash equilibrium of quantized game to that of its original continuous game was proved \cite{hao18b}.

\subsection{Bayesian Games}

Bayesian game is also called incomplete information game, where each player has several types to be assigned. A typical method to deal with Bayesian game is to convert it into a complete information game by replacing each player's payoff function by its expected value. Using STP, \cite{cheprb} provides a systematic method to convert different kinds of Bayesian games into different complete information games according to designed goals.

A new transformation that converts a Bayesian game into a so-called ex-ante agent game is proposed in \cite{wupr}. This transformation has several interesting properties, such as preserves potentiality. Moreover, using STP, the potential equation is designed.

\subsection{STP in Cooperative Games}

The previous results are all for non-cooperative games. In fact, STP also has some applications to cooperative games.

For instance, a set of unanimity games becomes a basis for finite cooperative games ${\cal G}^c(n)$. \cite{che13} proposed a matrix form for the characteristic functions of the unanimity games. In addition, a recursive algorithm to construct the matrix expression is also obtained.

It is well known that calculating the Shapley value is a heavy job. \cite{wan19} provided a recursive algorithm to calculate the transfer matrix $M_{\phi}$.

A recent development is: \cite{li21} proposed a new way to express the Shapley value. Using STP, a more convenient algorithm is obtained.

Banzhaf value is a valuable evaluation for the contribution of individuals in a cooperative game. Using STP, \cite{xia20} presented a simple algorithm to calculate the Banzhaf  value. In addition, it has been applied to diagnosing  genetic diseases.

\subsection{Game-based Control Technique}

Using STP, some game-based control technique has been developed. This is a new and promising direction.

A linear dynamic system-based evolutionary game, called a linear dynamic game, was proposed and investigated in \cite{wu17}. By resorting to STP, it was proved that the linear dynamic game is a PG. Algorithms have been developed for searching Nash equilibria.

A game-theoretic approach to a service chain routing in network function visualization with capacity limitation and user minimum target rate constraint was formulated into a potential game in \cite{le20,le21}.  It has potential engineering applications.

\subsection{Intelligent Games}

Using STP, some intelligent games can be formulated as NEGs. Then the results obtained for NEGs are applicable to these games. For instance, the problem of crossing river with wolf, sheep and vegetables and the problem with cannibals and missionaries were formulated as evolutionary games and then solved \cite{zha18}. A puzzle in the Tomb Raider - Anniversary was formulated as a formation control problem \cite{zha13}.

\subsection{Game-Theoretic Control}

An  `Hourglass' architecture of game-theoretic control, described in Fig.\ref{Fig.3.1},  was introduced by \cite{gop11}. We explain it as follows: assume a networked multi-agent system is given by a graph $(N,E)$. Our purpose is to maximize an overall performance criterion
\begin{align}\label{3.100.1}
\max_{x_i}J=F(x_1,x_2,\cdots,x_n),
\end{align}
where $x_i$ is the action of agent $i$, $i=1,2,\cdots,n$.

The basic idea of game-theoretic control approach to solve this optimization problem is to convert the overall system into a PG, then let the game converge to the maximum Nash equilibrium. Roughly speaking, the procedure is as follows:

\begin{itemize}
\item Step 1: Take $J$ as a potential function:
$$
P(x_1,x_2,\cdots,x_n):=J(x_1,x_2,\cdots,x_n).
$$
\item Step 2: Design local information based utility function $c_i$ for each agent $i$, such that the overall system becomes a potential game with $P=J$ as its potential function.

\item Step 3: Design a learning algorithm which provides SURs for all players. Such that using this algorithm, when each player trying to maximize its own utility function $c_i$, the overall system can reach its maximum value for $P=J$.

\end{itemize}

\vskip 5mm

\begin{figure}
\centering
\setlength{\unitlength}{5 mm}
\begin{picture}(8,6)\thicklines
\put(2,2){\framebox(4,1.5){$Potential\atop Game$}}
\put(0,0){\line(1,0){8}}
\put(0,0){\line(1,1){2}}
\put(8,0){\line(-1,1){2}}
\put(0,5.5){\line(1,0){8}}
\put(0,5.5){\line(1,-1){2}}
\put(8,5.5){\line(-1,-1){2}}
\put(2.5,1){$Algorithm\atop Design$}
\put(1.5,4.5){$Utility~Function\atop Design$}
\end{picture}
\caption{`Hourglass' Architecture of Game-Theoretic Control\label{Fig.3.1}}
\end{figure}
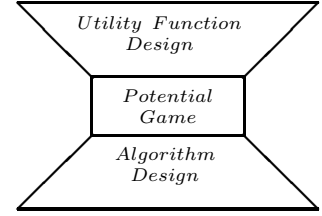

\vskip 5mm

STP approach has been used to solve the two design problems for game-theoretic control: utility function design and algorithm design.

\begin{itemize}
\item Considering a facility-based system, \cite{hao18b} provides a necessary and sufficient condition for design local information based utility functions respecting pre-assigned object function $J$.
\item For a class of evolutionary games, where the strategy profile transformation depends on the state evolution, \cite{li20} designed an algorithm, which updates each player's strategy using two step historic knowledge. The algorithm assures the convergence to maximum $J$.
\item For state-depending evolutionary games \cite{liu19} provides an overall framework for both utility and algorithm designs of game-theoretic control.
\end{itemize}

\section{Concluding Remarks}

This paper provides a comprehensive introduction for the STP approach to finite games, including non-cooperative games and cooperative games. The paper consists mainly of two parts: Section 2 explains why STP is a proper tool for formulating and investigating finite games. Section 3 consists of most game related topics investigated via STP.

Though the topics investigated by STP are various, there are still various problems that have not been touched, because game theory is a wide field. Moreover, game theory and control theory share many common issues: similar purpose, complementary techniques, etc., the cross discipline between them is a very prominent direction in both research and applications.

In the introduction we have discussed briefly the impact of control theory and game theory to AI. It is well known that logical system theory is also important for AI, while STP plays a fundamental role in the control of logical systems, in addition to its application to game theory, it is anticipated that STP could be a powerful tool in the development of AI.

\vskip 5mm

{\bf ACKNOWLEDGMENT}:  The authors thank Prof. Jun-e Feng, Prof. Jianquan Lu, Prof. Jiandong Zhu, Prof. Min Meng and Dr. Xiao Zhang for their valuable comments on the manuscript.

\end{document}